\newcommand{\ltappeq}{\raisebox{-0.6ex}{$\,\stackrel
{\raisebox{-.2ex}{$\textstyle <$}}{\sim}\,$}}
\newcommand{\gtappeq}{\raisebox{-0.6ex}{$\,\stackrel
{\raisebox{-.2ex}{$\textstyle >$}}{\sim}\,$}}

\documentclass[12pt,preprint]{aastex}
\usepackage{psfig}

\slugcomment{accepted for publication in ApJ}

\shorttitle{Far-UV Survey of 47 Tuc}
\shortauthors{Knigge et al.}

\begin{document}


\title{A Far-Ultraviolet Survey of 47~Tucanae. I. Imaging 
\footnote{Based on observations made with the
NASA/ESA Hubble Space Telescope, obtained at the Space Telescope
Science Institute, which is operated by the Association of
Universities for Research in Astronomy, Inc., under NASA contract NAS
5-26555. These observations are associated with proposal \#8219.}}


\author{Christian Knigge}
\affil{Department of Physics and Astronomy. University of Southampton, 
Southampton~SO17~1BJ,~UK}

\author{David. R. Zurek, Michael M. Shara}
\affil{Department of Astrophysics, American Museum of Natural History, New York, 
NY 10024}

\and

\author{Knox S. Long}
\affil{Space Telescope Science Institute, Baltimore, MD 21218}



\begin{abstract}
We present results from the imaging portion of a far-ultraviolet (FUV)
survey of the core of 47~Tucanae. We have detected 767 FUV sources,
527 of which have optical counterparts in archival HST/WFPC2 images of
the same field. 

Most of our FUV sources are main-sequence (MS) turn-off
stars near the detection limit of our survey. However, the FUV/optical
color-magnitude diagram (CMD) also reveals 19 blue stragglers (BSs), 17
white dwarfs (WDs) and 16 cataclysmic variable
(CV) candidates. The BSs lie on the extended cluster MS, and four of
them are variable in the FUV data. The WDs occupy the top of
the cluster cooling sequence, down to an effective temperature of
$T_{eff} \simeq 20,000$~K. Our FUV source catalog probably contains
many additional, cooler WDs without optical counterparts. Finally, the
CV candidates are objects between the WD cooling track and the
extended cluster MS.

Four of the CV candidates are previously known or suspected
cataclysmics. All of these are bright and variable in the FUV. Another 
CV candidate is associated with the semi-detached binary system V36
that was recently found by Albrow et al. (2001). V36 has an orbital 
period of 0.4 or 0.8~days, blue optical colors and is located within
1~arcsec of a Chandra x-ray source. A few of the remaining CV candidates
may represent chance superpositions or SMC interlopers, but at least
half are expected to be real cluster members with peculiar
colors. However, only a few of these CV candidates are possible 
counterparts to Chandra x-ray sources. Thus it is not yet clear which, if
any, of them are true CVs, rather than non-interacting MS/WD binaries
or Helium WDs.



\end{abstract}


\keywords{globular clusters: individual: name: 47~Tucanae; stars:
novae, cataclysmic variables; stars: blue stragglers; stars: white
dwarfs; ultraviolet: general}


\section{Introduction}

Globular clusters (GCs) are fantastic stellar crash test
laboratories. Violent encounters between binaries and single stars in
dense cluster cores 
give rise to exotic stellar populations, such as blue stragglers
(BSs), cataclysmic variables (CVs), low-mass x-ray binaries and
milli-second pulsars. All of these objects have 
considerably bluer spectral energy distributions than the
non-degenerate stars that make up the bulk of the GC
mass. Observations at short wavelengths are thus ideally suited to the 
detection and characterisation of these relatively rare stellar
species. This is nicely illustrated by the recent Chandra survey of 
47~Tuc (Grindlay et al. 2001a), which finally showed that the
theoretically predicted, extensive population of interacting close 
binaries really does exist there. Nevertheless, important puzzles
still remain. For example, even though Grindlay et
al. (2001a) estimated that there are 30 CVs with $L_x \gtappeq
10^{30}$~ergs~s$^{-1}$~in 47~Tuc --  far more than found in any
previous study -- this is at best 1/3 of the number predicted by 
tidal capture theory (Di Stefano \& Rappaport 1994).

Given the striking differences between optical and Chandra x-ray 
images of 47 Tuc, it seems clear that observations at intermediate
wavelengths -- i.e. the far-ultraviolet (FUV) -- would be invaluable 
in identifying and characterising both the Chandra sources and other
exotic objects in the cluster. For example, it is well known that CVs
and young white dwarfs (WDs) radiate much of their luminosity in the
FUV waveband. We have therefore embarked on a program to study 
the cores of nearby GCs at FUV wavelengths, using both imaging and
slitless spectroscopy. The extraordinary spatial resolution and
FUV sensitivity of the Space Telescope Imaging Spectrograph (STIS)
aboard the Hubble Space Telescope (HST) make this the instrument of
choice for our study. Here, we  present first results from a campaign
on 47~Tuc, which clearly demonstrate the value of FUV observations to
the study of GCs.

\section{Observations and Data Reduction}

\subsection{FUV Photometry}

Thirty orbits of STIS/HST observations of 47 Tuc have been obtained,
comprised of six epochs of five orbits each (HST program GO-8219). In
each epoch, we carried out FUV imaging and slitless spectroscopy. 
Consecutive observing epochs were as closely  
spaced as a few days and as widely separated as a year. Typical
exposure times in the FUV were 600~s. Our program is therefore
sensitive to variability on time-scales ranging from minutes to
months. In this first paper arising from our survey, we will focus on
the results obtained from the FUV imaging aspect of the program;
the analysis of the slitless spectroscopy is in progress and will be
presented elsewhere. 

All of our FUV observations (both imaging and spectroscopic) used the
FUV-MAMA detectors and were taken through the F25QTZ filter. The
purpose of this filter is to block geocoronal Ly$\alpha$, OI
1304~\AA~and OI] 1356~\AA~emission which would otherwise produce a high 
background across the detector in our slitless spectroscopy. The
effective bandpass with this instrumental set-up is
1450~\AA~--~1800~\AA. The 1024$\times$1024 pixel FUV-MAMA detector
covers approximately 
25\arcsec$\times$25\arcsec, at a spatial resolution of about
0.043\arcsec~(FWHM). Our field of view (FoV) was chosen to overlap with
archival HST observations of 47~Tuc and included the cluster center.

Source detection and photometry was carried out on a deep FUV image 
that was constructed by registering and  co-adding all of the individual
FUV imaging exposures. We note that -- like x-ray detectors,
but unlike optical CCDs -- the FUV/MAMAs are photon counting devices 
with no read noise and very little dark current (approximately
$7\times10^{-6}$~c~s$^{-1}$~pix$^{-1}$). As a result, sky and faint 
source pixels tend to show either exactly zero or one count in a 
typical FUV exposure. This implies that median filtering is a
sub-optimal way of combining multiple FUV exposures. Instead, FUV
images should be combined by direct summing or averaging.  

We obtained two FUV imaging exposures at the beginning and end of each
observing epoch, for a total of 24 FUV images. The registration
and co-addition of these images was performed with routines available
in the {\sc iraf/images}
\footnote{IRAF (Image Reduction and Analysis Facility) is distributed
by the National Astronomy Optical Observatories, which are operated by
AURA, Inc, under cooperative agreement with the National Science
Foundation.}
package. In doing so, we took care to mask
known image defects (e.g. hot pixels) by inspecting the data-quality
file associated with each individual exposure. The final, deep FUV image
corresponds to a total exposure time of approximately 14,600~s.

Source detection and photometry on the co-added FUV image was carried 
out with the SExtractor software (Bertin \& Arnout 1996). This 
was preferred to other packages because our FUV images share
characteristics with both standard CCD images (e.g. compact PSF) and
x-ray images (e.g. low background, complex PSF shape). The software of
choice has to work well with both types of characteristics, and the
SExtractor fits this bill: it was designed primarily for use with CCD
images, but is also known to work well on x-ray images (Valtchanov,
Pierre \& Gastaud 2001). 

The SExtractor uses a matched filtering method for source
detection. When dealing with CCD images, it is usually sufficient 
to use a simple analytic filter (e.g. a Gaussian) whose width is
chosen to match that of the detector point spread function
(PSF). However, the FUV/MAMA PSF is quite complex and asymmetric. We
therefore designed our own matched filter from an empirical PSF that
was constructed using the {\sc daophot} package within {\sc iraf}.

The detection threshold within SExtractor was set to 5$\sigma$ above
the local background, with the minimum source area set to 5
pixels. However, it is important to note that these numbers refer to
the {\em filtered} image. A more transparent way to characterise our
detection threshold is to consider the completeness of our
source detection procedure as a function of magnitude. As described in 
Section~2, artificial star tests suggest that our FUV survey is almost 
complete to $m_{FUV} \simeq 24$. Here, $m_{FUV}$ is the FUV
magnitude of a source in the STMAG system, defined by 
\begin{equation}
m_{FUV} = -2.5 \log{F_\lambda} - 21.1,
\end{equation}
where $F_\lambda$ is the flux of a hypothetical flat-spectrum
source. Thus $m_{FUV} \simeq 24$ corresponds to $F_{\lambda} \simeq 9
\times 10^{-19}$~ergs~cm$^{-2}$~s$^{1}$~\AA$^{-1}$. In terms of
recorded counts, we have 
\begin{equation}
F_\lambda = \frac{{\rm Counts} \times {\rm PHOTFLAM}}{{\rm Exposure~
Time}},
\end{equation}
where PHOTFLAM ($= 1.04 \times
10^{-16}$~ergs~cm$^{-2}$~\AA$^{-1}$~counts$^{-1}$ for our set-up) 
is the inverse sensitivity of the detector/filter
combination. Given the effective
exposure time of the co-added FUV frame, our $m_{FUV} \simeq 24$
completeness threshold corresponds to approximately 125
source counts; about 80 of these fall within our photometric
aperture (see below). Thus our detection threshold corresponds to $S/N
\simeq 9$ at the completeness limit. As a last step, we rejected
sources that were too close to the image edges, resulting in a final
catalog of 767 FUV sources.

Since the FUV image is not particularly crowded, we were able to use
simple aperture photometry to measure the magnitudes of the detected
FUV sources. All FUV magnitudes quoted below are on the STMAG system
and were calculated using an aperture with a radius of 5 pixels. The
aperture correction to infinite radius was estimated using both empirical and 
theoretical PSFs. The theoretical PSF was calculated with the TinyTim 
software\footnote{TinyTim is available at
{\tt http://www.stsci.edu/software/tinytim}.}. 

We also carried out a search for variability among our FUV
sources. This was done by registering individual FUV frames to the
co-added frame and performing aperture photometry at the positions of
the previously detected FUV sources on these frames. Likely variables
were identified by calculating the standard deviation, $\sigma$, of
each light curve and searching for outliers in a plot of $\sigma$ vs
$m_{FUV}$. Candidate variables were then inspected by eye on the
images and removed from the list if their variability appeared to be
spurious (e.g. due to image defects or a location near the image
edges). This procedure finally allowed us to identify 8 FUV sources
that appear to be genuinely variable. These will be discussed in more
detail in Section~\ref{results}.

\subsection{Optical Photometry}
\label{optphot}

In order to find optical counterparts for our FUV sources, we used the
HST archive to construct a deep U-band image of the STIS FoV. More
specifically, we extracted from the HST archive all  
WFPC2/PC/F336W ($\simeq$ U) exposures that contained at least part of
our FoV. We registered and combined these images, and carried out
source detection and PSF-fitting photometry on the co-added frame
using {\sc daophot} (Stetson 1987). Magnitudes were
again placed on the STMAG system, but it is important to note that
stars brighter than $m_{336} \simeq 16$ were generally saturated in
the co-added frame. At the other extreme, the optical photometry is
not sensitive to objects fainter than $m_{336} \simeq 22 - 23$.

\subsection{Matching FUV and Optical Catalogs}
\label{chance}

We also extracted and combined all WFPC2/PC/F218W images covering our
FoV from the HST archive. Even for the bluest objects (hot WDs), the
co-added F218W image does not go as deep as the co-added F336W
image. However, the F218W bandpass lies between the STIS/FUV/F25QTZ  
and WFPC2/F336W bandpasses, making the F218W frame a useful
intermediary in the process of matching up FUV and F336W
sources. Ferraro et al.'s (2001) list of BSs,  
hot WDs and UV-excess stars in the core of 47~Tuc provided a
particularly convenient starting point in this context. These authors
provide F218W positions for all objects on their list, many of
which turn out to be detectable in both FUV and F336W. 
Starting with this initial list of matches, we iteratively improved
the transformation between F336W and FUV positions by finding
additional matches and rederiving transformation coefficients. With 
our final transformation, there are 4894 U-band sources within our
effective FUV FoV.

In our search for optical counterparts, we adopted a maximum
difference of 1.5 STIS/FUV-MAMA pixels ($\simeq 0.038\arcsec \simeq
0.8$~WFPC2/PC pixels) between 
the centers of FUV sources and their F336W counterparts. Even with
this conservative 
criterion, we managed to find counterparts to 527 FUV sources. We chose
such a restrictive matching radius in order to keep the likely number
of false matches at an acceptable level. This is an important
consideration, since with 767 FUV and 4894 optical sources sharing
approximately 1~million pixels, we certainly expect some false matches
to occur. The expected number of false matches depends on the numbers of
objects in the two source lists, the matching tolerance and the actual
number of objects matched. Based on these parameters, we estimate that
our list of optical counterparts contains approximately 8 false 
matches overall, but only 3 among FUV sources brighter than $m_{FUV} =
22$. This distinction is important because most of the interesting
objects in our CMD are brighter than this limit.

\section{Results and Discussion}
\label{results}

\subsection{Comparison of FUV, Optical and X-ray Imaging}

In Figure~1, we show a comparison of the F336W and FUV images of the
same 25\arcsec$\times$25\arcsec~field near the core of 47~Tuc. The
optical image is vastly more crowded, because the majority of
main-sequence stars, red giants and horizontal branch stars are too
cool to show up in the FUV exposure. We have overlayed onto the FUV
image the predicted positions of known, blue sources from Geffert,
Auri\'{e}re \& Koch-Miramond (1997) and also the predicted positions of the Chandra x-ray
sources from Grindlay et al. (2001a). In both cases, the positions were
determined by first placing the sources on the F336W frame, using the 
{\sc invmetric} task within {\sc iraf/stsdas}, then improving these
positions by using three sources that are common to both lists and are
also detected in both our F336W and FUV frames. These three sources
are AKO9 [W36], V1 [W42, X9] and V2 [W30, X19] (the identifications in
square brackets refer to alternative designations in the literature;
'W' numbers, in particular, refer to Table~1 of Grindlay et
al. 2001a; `X' numbers refer to Table~2 of Verbunt \& Hasinger
1998). We finally transformed the positions to the STIS/FUV frame,
using our previously derived transformation. 

Figure~1 shows that almost all of the blue sources listed by Geffert
et al. (1997) have certain or likely FUV counterparts. This was to be
expected, but is nevertheless 
important: it provides an external check on our astrometry and
supports our contention that FUV imaging is an excellent way of
finding interesting GC sources. We find fewer obvious FUV counterparts 
to the Grindlay et al. (2001a) x-ray sources within our FoV. This is not so surprising, given that 
approximately 70\% of their sources are expected to be very faint FUV
sources (millisecond pulsars and x-ray active main sequence
binaries). A more interesting question is whether we detect those
Chandra sources that are definitely expected to be FUV 
bright. There are only four such sources in our FoV, all of
which were classified by Grindlay et al. as CVs. Three of these are 
thought to be associated with previously
known or suspected CVs, namely our reference stars AKO~9, V1 and V2;
all three are also clearly detected in our FUV images. The fourth source --
denoted W15 by Grindlay et al. (2001a) -- is a CV candidate that was not
known prior to the Chandra survey of 47~Tuc. It, too, has a likely
counterpart in our FUV images (see Figure~1 and
Section~\ref{cvs}). 

\subsection{The FUV luminosity function}

In Figure~2, we show the FUV luminosity function (LF) for the 767 stars 
we detected in our FUV image. The main feature of the LF is the sharp
rise towards $m_{FUV} \simeq 24$. As we will show below, this feature 
is due to main-sequence turn-off (MSTO) stars which become
detectable near this magnitude. A second interesting aspect of the LF
is that the brightest source in the sample earns that distinction by a
considerable margin (approximately 1.75~mags). This source is AKO~9,
which we will return to in Section~\ref{cvs} below. 

In order to assess the completeness of our sample, we used the {\sc
addstar} routine within {\sc daophot} to carry out 
artificial star tests. The results 
are shown in the top panel of Figure~2 and suggest that our survey is
close to complete to roughly $m_{FUV} \simeq 24$. However, it is
worth noting that (i) both {\sc addstar} and SExtractor used the same
PSF, and (ii) the {\sc addstar} routine does not account for the
quantization of counts at low count levels. As a result, our
completeness estimates could be slightly optimistic. 

\subsection{The FUV/Optical Color-Magnitude Diagram}

In Figure~3, we present the FUV-optical color-magnitude diagram (CMD)
of  47~Tuc. Several distinct stellar populations are present in CMD,
among them WDs, BSs, MSTO stars and, last but not least, CVs. We have
also calculated and plotted a set of theoretical tracks whose purpose is to
aid in the interpretation of the CMD. We will first describe the
calculation of these tracks and then discuss the implications of our
CMD for our understanding of the stellar populations within it. 

\subsubsection{Synthetic Photometry}

The starting point for our synthetic photometry was the recent
work by VandenBerg (2000), who modelled the optical CMD of 47~Tuc. Don
VandenBerg kindly provided us with the corresponding set of isochrones
and zero-age horizontal branch (ZAHB) models, from which we extracted
the stellar parameters of main sequence stars, sub-giants, red giants
and horizontal branch stars in 47~Tuc. We then used {\sc synphot}
within {\sc iraf/stsdas} to calculate the FUV and optical magnitudes
of stars on these sequences. This was achieved by interpolating 
on the Kurucz grid of model stellar atmospheres and folding the
resulting synthetic spectra through the response of the appropriate
filter+detector combinations. The cluster parameters we adopted in our 
calculations are those of VandenBerg (2000), i.e. $E(B-V) = 0.032$,
$d = 4510$~pc, ${\rm [Fe/H]} = -0.83$ and ${\rm[\alpha/Fe]} = 0.3$. 
VandenBerg's study puts 47~Tuc's age at approximately 11.5~Gyr,
so we calculated isochrones for 10, 12 and 14~Gyr. 

We also calculated an approximate BS sequence for our CMD. For this,
we assumed BSs to lie near the zero-age main sequence (ZAMS) of the
cluster, and used the fitting formulae of Tout et al. (1996) to
estimate the appropriate stellar parameters. The corresponding FUV and
optical colors were again estimated from the Kurucz model grid
within {\sc synphot}. The same technique was 
used to estimate the location of the ZAMS of the
Small Magellanic Cloud (SMC), which is relevant because 47~Tuc is
located in front of the SMC. For the SMC, we adopted a metallicity of
${\rm [Fe/H]} = -0.68$, a distance of 58.6~kpc and the same reddening
as for 47~Tuc (B\"{o}hm-Vitense 1997). We also 
calculated a theoretical WD cooling sequence for our CMD. In
doing so, we adopted a mean cluster WD mass of 0.53~M$_{\odot}$
(Renzini \& Fusi-Pecci 1988) and used DA WD model atmospheres to
calculate the predicted colors. For models with $T_{eff} \leq
15,000$~K, we used a small grid of model atmospheres calculated with 
{\sc tlusty/synspec} (Hubeny, Lanz \& Jeffery 1994) and kindly
provided to us by Ivan Hubeny. For models with higher $T_{eff}$, we
used the Vennes et al. (1993) grid of DA WD models, which was kindly
provided to us by Stephane Vennes.

\subsubsection{Main-sequence stars, sub-giants, red-giants and
horizontal branch stars}

Returning to Figure~3, we first note the prominent clump of stars in the
upper right quadrant. This clump contains most of the stars in the CMD
and happily coincides almost exactly with the expected location of the
MSTO. 

We next focus on the smaller clump of stars near the predicted 
location of the ZAHB (i.e. above and to the blue of the MSTO). This 
second clump (at $m_{F336W} \ltappeq 16$) contains mostly 
saturated horizontal branch stars (c.f. Section~\ref{optphot}); i.e. 
its apparent location below the ZAHB track is an artifact of saturation. 

Finally, there is also a trail of stars between these two clumps, which 
cannot be explained as an artifact due to saturation. We have therefore 
used additional archival data to investigate the nature of the sources
on this trail and find that they are red giants. For comparison, the
synthetic red giant branch is approximately 4-5 magnitudes redward of
this trail. 

In order to understand this apparent discrepancy, it is important to
realize  
that our CMD does not imply that these stars are FUV bright, only that
they are not as FUV faint as stellar atmosphere models predict. This
is not particularly surprising. After all, the Kurucz model
atmospheres include only the {\em photospheric} stellar flux. However,
the FUV waveband is well down the Wien tail of a red 
giant's photospheric spectrum, so that {\em chromospheric} emission can 
totally dominate the FUV output of these systems. Figure~1 of
Robinson, Carpenter \& Brown (1998) nicely illustrates this effect 
for two isolated K Giants. We finally note that the
chromospherically-dominated FUV spectra of red giants are rich in
emission lines; these might be detectable in our slitless
spectroscopy.

\subsubsection{Blue Stragglers}

The BS sequence can be seen as a trail of stars starting at the MSTO
and going upwards and to the left. The slight discrepancy between the
observed BS sequence and the synthetic one is not surprising, since
the latter assumes that the BSs lie on the ZAMS. In reality, BSs may
be expected to be somewhat evolved, explaining their location above
and to the red of the ZAMS. In total, there are 19 BSs in our CMD,
extending all the way from the MSTO (at approximately 0.9~M$_{\odot}$)
to perhaps 1.5~M$_{\odot}$ (judging from the ZAMS models). Four of
these (circled in Figure~3) appear to be variable in our FUV
photometry. This is plausible, given that variability is a known
phenomenon among BSs (e.g. Gilliland et al. 1995).  

\subsubsection{White Dwarfs}
\label{wd}

The WD cooling sequence is clearly visible in the bottom left quadrant
of the CMD. The faintest WDs on this sequence have $T_{eff} \simeq
20,000$~K, but it is worth noting that this detection limit is imposed
by the depth of the optical data. Indeed, our FUV detection limit of
$m_{FUV} \simeq 24$ corresponds to a WD with $T_{eff} \simeq
11,500$~K. 

It is interesting to ask whether the numbers of WDs we detect are in
line with expectations. A simple estimate of the 
expected number can be made following Richer et al. (1997). The number
of stars in two post-main-sequence phases is, in general, proportional
to the duration of these phases. We will use horizontal branch (HB)
stars as a reference point, and take $N_{HB}$ as the number of HB
stars in our FoV and $\tau_{HB} \simeq 10^8$~yrs
as their lifetime (e.g. Dorman 1992). We then expect to find
approximately 
\begin{equation}
N_{WD} (T_{eff} > T_{lim}) \simeq N_{HB} \times
\left[\frac{\tau_{WD}(T_{lim})}{\tau_{HB}}\right],
\end{equation}
\label{N}
WDs hotter than $T_{lim}$ in the FoV, where
$\tau_{WD}(T_{lim})$ is the time it takes a WD to cool to
$T_{lim}$. Inspection of published optical observations shows that
$N_{HB} \simeq 50$ (Guhathakurta et al. 1992). We note that the 
estimate in Equation~\ref{N} ignores the effect of mass
segregation. This is a reasonable first approximation in our case,
since the cooling time-scales among our WD population are short and 
comparable to the core relaxation time of $\sim 10^8$~yrs [Harris
1996]). 

For a first check, we will assume that the 17 WDs in our CMD represent
the complete WD population in our FoV down to $T_{eff} \simeq
20,000$~K. The true number is likely to be higher, since 
our optical data is likely to be seriously incomplete at the faint end
of the WD sequence. If we nevertheless adopt this assumption 
and use the models of Althaus \& Benvenuto (1998) to estimate the 
corresponding cooling time, we find $\tau_{WD} \simeq 5 \times
10^7$~yrs and hence $N_{WD}(T_{eff} > 20,000{\rm ~K}) \simeq 25$. As
expected, this is slightly larger than the observed number, but 
in the right ballpark. A second estimate can be made by assuming that
all FUV sources with $m_{FUV} < 22$ (corresponding to $T_{eff} \simeq 
14,500$~K) are WDs, unless their location in the CMD contradicts
this. This allows us to include sources without optical counterparts, 
on the assumption that all such sources are optically faint WDs. There
are a total of 121 FUV sources satisfying $m_{FUV} < 22$ in our
catalog, 34 of which are clearly not single WDs (based on their
location in the CMD). We therefore estimate
that the number of WDs with $T_{eff} > 14,500{\rm ~K}$ is 87. The true
number is probably somewhat lower, since some of the FUV sources
without optical counterparts may, for example, be faint CVs (see
below). The cooling time for this temperature is roughly $1.8 \times
10^{8}$~yrs and hence the predicted number of WDs is $N_{WD}(T_{eff} >
14,500{\rm ~K}) \simeq 90$. Observed and predicted numbers appear to
be in reasonable agreement.

\subsubsection{Cataclysmic Variables}
\label{cvs}

We finally turn to the CV candidates that are revealed by our 
CMD. Given that CVs are binary systems containing an accreting WD and
(usually) a MS star, we expect them to be located between the WD
cooling sequence on one side and the MS (and its BS extension) on the
other. We will refer to this area of the CMD as the ``CV zone''. There
are 16 objects in the CV zone, not including the star just off the
lower left of the MSTO. How many of these sources are likely to be
real CVs? 
   
We first note that four objects in the CV zone -- AKO~9, V1, V2 and
W15 -- are already confirmed CVs. All four of these sources are FUV bright 
and variable, and all four are Chandra x-ray sources that were also
classified as CVs by Grindlay et al. (2001a). In the case of AKO~9 and
V2, our slitless spectroscopy has already provided spectroscopic
confirmation as well, via the detection of FUV emission lines (Knigge
et al., in preparation; see Knigge et al. 2002 for a preliminary
report). The identification of AKO~9 as a CV is especially important,
since the nature of this 1.1-day binary had been unclear since its
discovery by Auri\'{e}re, Koch-Miramond \& Ortolani (1989). A more
detailed description of our FUV data on this source is in
preparation.

Another one of our CV candidates also deserves special mention. This
is the source labelled as V36 in both Figures~1 and 3. This 
identification follows the notation of Albrow et al. (2001), who show 
that V36 is likely to be a semi-detached binary system with an orbital
period of 
0.4 or 0.8 days. V36 was previously classified as a BS by Paresce et
al. (1991) and has also been noted as
an unusually blue source by De Marchi, Paresce \& Ferraro (1993),
Geffert et al. (1997) and Ferraro et al. (2001).\footnote{See the
discussion in Ferraro et al. (2001) for the names assigned to V36 
in the various catalogs.} Based on the FUV imaging and astrometry
shown in Figure~1, we now find that V36 is also a strong FUV emitter 
and located within 1~arcsec of a Chandra x-ray source. We therefore
suggest that V36 is a strong CV candidate (see, however, the {\em note
added} at the end of this paper).

Turning to the remaining 11 sources in the CV zone, we first consider
the possibility that some of them are due to chance superpositions of
physically unrelated WDs and MS stars. All of the sources are 
relatively FUV bright ($m_{FUV} \ltappeq 22$) and therefore
located in a region of the CMD where we should expect only about 3 false
matches (c.f. Section~\ref{chance}). The resulting spurious sources
would most likely be located near the peak of the optical luminosity
function, i.e. around $m_{336} \simeq 18$ (e.g. Howell et
al. 2001). Thus any false matches amongst the objects in the CV zone
are most likely to come from the 7 sources with $m_{336} < 20$ and
$m_{FUV} - m_{336} > 0$. However, only three of these -- the three CV
candidates closest to V36 in the CMD -- show an offset in excess of 1
pixel between FUV and optical positions.

A second point to consider is whether some of the objects in the 
CV zone may be background stars in the SMC halo behind 47~Tuc. The
expected location of the SMC ZAMS is indicated on the CMD. However, it
is important to note that the bulk 
of SMC stars in this region follows an isochrone whose MSTO lies at 
about $m_{336} \sim 22.5$ (c.f. Zoccali et al. 2001). The likely
degree of SMC contamination in our data can be quantified by
comparison with the HST/WFPC2 observations of Zoccali et
al. (2001). Their FoV covers approximately
17,000~arcsec$^2$~centered 6.5\arcmin~west of the cluster
center. Within this field, there are perhaps 10 possible SMC
interlopers brighter than $m_{336} \sim 22.5$ (see their Figure~1, but
note that their quoted magnitudes are instrumental). Scaling this
number to the area covered by our roughly 630~arcsec$^2$ FoV, 
we may expect to find about 0.4 SMC interlopers in our
CMD. Even allowing for small number statistics, it is unlikely that
there are more than one or two SMC stars in the CV zone. 
 
The third question we must ask is whether some of the objects in the
CV zone are affected by blending, image defects, edge effects, etc 
in the FUV image. We have therefore inspected by eye the positions of
all 16 objects in the co-added FUV image (c.f. Figure~1). Based on
this, the two sources closest to the BS sequence within the CV zone
deserve further comment. One of them lies quite close to an image
edge, but our aperture
photometry for it should not be compromised significantly. In any case,
we would expect edge effects to {\em reduce} the apparent brightness
of this source, i.e. any correction would probably make this object
appear even bluer and thus move it further into the CV zone. The other 
object lies within a few pixels of a fainter FUV source. This may have
led to a mild overestimate of its FUV brightness, but the shift
required to bring it close to the MSTO (about 2 mags) appears to be
much larger than any implied correction.  

Fourth, we have to look more carefully at our definition of the ``CV
zone''. More specifically, we must ask if the colors of 
the objects in this zone are actually consistent with those of CVs. We
have addressed this question by plotting the locations of several 
proto-typical field CVs in our CMD, after placing these systems at the
distance and reddening of 47~Tuc. The FUV magnitudes of
the field CVs were calculated by running their IUE spectra through
{\sc synphot}; their F336W colors were estimated from U-band magnitudes
in the literature (primarily Bruch \& Engel 1994). Some field CVs
appear twice in Figure~3, representing the same object in different
states (e.g. dwarf novae in quiescence and outburst). We stress that
this comparison between cluster CV candidates and field CVs is fraught
with uncertainties, since the stellar environments and parameters of
the two types of CVs are quite different. Nevertheless, it is clear
from Figure~3 that most of the cluster sources in the CV zone of the
CMD lie reasonably close to the positions of field CVs. The comparison
also shows that the spectrum of some field CVs, like AM Her, can at
times be dominated by their WDs, in both FUV and U/F336W
bandpasses. This means that a few of the FUV sources on the WD cooling
track may yet turn out to be CVs. 

Fifth and finally, we must look at other types of stars that may be 
found in the CV zone, such as non-interacting MS/WD binaries. Indeed,
given that there may be $\simeq 90$ young 
WDs down to $m_{FUV} \simeq 22$ (Section~\ref{wd}) and that the binary
frequency in the core of 47~Tuc is about 15\% (Albrow et al. 2001),
{\em all} of the unconfirmed CV candidates could be non-interacting
MS/WD binaries. Another stellar type that may occupy the CV zone are
Helium WDs, which have, for example, been found in NGC~6397 (Cool et
al. 1998). Both MS/WD binaries and He WDs would be consistent with the
apparent lack of FUV variability amongst our unconfirmed CV
candidates.

Based on the discussion above, we conclude that (i) a conservative 
estimate for the number of ``real'' sources in the CV zone -- as
opposed to SMC interlopers, image defects and products of chance
coincidences -- is $N_{real} \simeq 10$; (ii) that the ``CV zone''  
is a reasonable way to identify CV candidates; and (iii) that
additional information will be needed to distinguish CVs from
non-interacting MS/WD binaries and He WDs in the CV zone. With these
conclusions in mind, we will consider both a best-case and a
worst-case scenario. In the best-case scenario, the number of CVs in 
our CMD is $N_{CV} \simeq  N_{real} \simeq 10$; in the worst-case
scenario, none of the currently unconfirmed CVs will turn 
out to be real cataclysmics and $N_{CV} \simeq 5$. How do these
numbers compare with theoretical expectations?

The simulations of Di Stefano \& Rappaport (1994) predict that 47~Tuc
should contain approximately 190 CVs that were formed by two-body
tidal capture events (Fabian, Pringle \& Rees 1975). Approximately
half of these are expected to be located within one core radius of the
cluster. Other CV formation channels exist 
(Davies 1997), but are probably less important than tidal capture in
the cluster core. Given that our effective FoV samples approximately 
35\% of the 47~Tuc's core ($r_{core} = 23$\arcsec; Howell et al. 2001),
our STIS image should contain roughly 30 CVs. However, many of these 
are predicted to be  extremely faint, post-minimum
period systems (``period-bouncers''), which we cannot hope to detect 
in our F336W data. This is confirmed by Figure~3, which shows that our 
F336W detection limit would not allow us to detect the most likely
``period-bouncer'' among field CVs -- WZ~Sge -- in its normal, quiescent
state. If we only count those CVs with predicted x-ray  luminosities
$L_{x} > 10^{31}$~erg~s$^{-1}$ in the model of Di Stefano \& Rappaport
(1994) -- a rough dividing  
line between pre- and post-bounce CVs (c.f. their Figures 3 and 4) 
-- the predicted number drops by roughly another factor of two. 
Thus the tidal capture models of Di Stefano \& Rappaport (1994) 
predict that our CMD should contain approximately 15 CVs. 

We conclude that if our optimistic estimate of $N_{CV} \simeq 10$ is
correct, the observed-to-predicted ratio is about 2/3. This is not
significantly different from unity, given the small number statistics
involved. Thus the observed number of cluster CVs may, for the first
time, be consistent with predictions based on tidal capture
theory. But even in our worst-case scenario of $N_{CV} \simeq 5$, the
observed-to-predicted ratio is still about 1/3. The same ratio was
found by Grindlay et al. (2001a), based on their Chandra observations
of the whole cluster.

We close this section by noting that only a few of our 11 unconfirmed
CV candidates are located close to known x-ray sources
(c.f. Figure~1). Assuming that this is not simply due to remaining  
uncertainties in the astrometry, there are two possible explanations
for this. First, perhaps the worst-case scenario is close to the
truth, and few or none  of our new CV candidates are actually CVs. 
Second, our FUV survey is turning up new, x-ray faint CVs. Given the
x-ray properties of field CVs (e.g. van Teeseling, Beuermann \&
Verbunt 1996) and the detection limit of the Chandra survey of 47~Tuc
($L_x \sim 10^{30}$~erg~s$^{-1}$), this would be somewhat
surprising. We may soon be able to decide between these possibilities
on the basis of our time-resolved, slitless FUV spectroscopy.

\section{Conclusions}

We have presented first results from the imaging part of a FUV survey
of 47~Tuc. The great benefit of moving to the FUV is that most
ordinary cluster members are too cool to show up in this
bandpass. This dramatically reduces crowding and makes it easy to find
hot and/or exotic objects such as CVs, BSs and young WDs. The same
lack of crowding even allows us to carry out {\em slitless},
multi-object FUV spectroscopy of the dense cluster core.

In this first analysis of our FUV observations of 47~Tuc, we have
focused on the imaging aspect of our survey. By matching our FUV 
source catalog to a list of sources in a deep WFPC2/F336W ($\simeq$~U)
image of the same field, we have been able to find 17 WDs, 19 BSs and
16 candidate CVs. All four previously known CVs in our FoV --
including one which was only recently found on the basis of Chandra
x-ray imaging (Grindlay et al. 2001a) -- are among these 16
candidates. Another one of our CV candidates is associated with the 
semi-detached binary system V36, which was recently
disovered by Albrow et al. (2001). V36 has an orbital period of 0.4 or
0.8~days, blue optical 
colors and is located within 1~arcsec of a Chandra x-ray source. A few 
of the 11 remaining CV candidates may represent chance superpositions or
SMC interlopers, but at least half are
expected to be real cluster members with peculiar colors. However,
only a few of these are located close to known x-ray sources. Our
existing FUV spectroscopy and/or additional HST/WFPC2 images in
different bandpasses should soon allow us test which, if any, of them
are true CVs.

In summary, this first, deep FUV survey of a GC core highlights the
great benefits of FUV observations to the study of rare and exotic
stellar populations, not just in GCs, but also in other environments.

{\em Note added:} The referee of this paper, Peter Edmonds, has
informed us that unpublished astrometry carried out by himself and Ron
Gilliland suggests a different identification for the Chandra source
close to V36. We await to see details of this work in the
literature.At the present time, we think even the remaining evidence
-- V36's  blue FUV/optical colors, its orbital period of 0.4~or 0.8~days and
Albrow et al.'s suggested classification of V36 as a semi-detached
system -- is sufficient to support our classification of V36 as a
strong CV candidate. We hope that our FUV spectroscopy will soon shed
additional light on the nature of this intriguing system. 

\acknowledgments We are grateful to the referee, Peter Edmonds for a
detailed report that helped us improve this paper. Support for
proposal \#8219 was provided by NASA 
through a grant from the Space Telescope Science Institute, which is
operated by the Association of Universities for Research in Astronomy,
Inc., under NASA contract NAS 5-26555.

\clearpage



\figcaption[both_images_150dpi.ps]{{\em Left Panel:} The
co-added FUV image of the core of 47~Tuc. The image is approximately
25\arcsec$\times$25\arcsec in size and includes the cluster center
(marked as a white cross; position taken from Guhathakurta et
al. 1992). For comparison, 47~Tuc's core radius is 23\arcsec (Howell,
Guhathakurta \& Gilliland 2001). The positions of known blue objects 
from Geffert et al. (1997; green squares), Chandra x-ray sources from
Grindlay et al. (2001; large yellow circles) and of the CV candidates
discussed in Section~\ref{cvs} (small blue circles) are marked. The
four confirmed CVs within our FoV are labelled with their
most common designations. The
image is displayed on a logarithmic intensity scale and with limited
dynamic range so as to bring out some of the fainter FUV sources. 
{\em Right Panel:} The co-added WFPC2/F336W image of the same
field. This image, too, is shown with a logarithmic intensity scale
and limited dynamic range. \label{fig1}}

\figcaption[lf1.eps]{{\em Bottom Panel:} The luminosity function
determined from our FUV source catalog. {\em Top Panel:} The
completeness of our catalog as a function of magnitude, as estimated
from artificial star tests; see text for details.\label{fig2}}

\figcaption[fuv_f336_cmd_new.ps]{The FUV/optical color-magnitude
diagram. The positions of FUV sources with optical counterparts are
shown as red dots. Variable FUV sources are additionally marked with
black circles, and the four confirmed CVs in our FoV are
labelled. The two short-dashed diagonal lines are lines of constant
FUV magnitude; they mark the completeness limit of our catalog
($m_{FUV} = 24$) and a rough dividing line between WDs, BSSs and CVs
on the one hand, and MSTO stars, HB stars and red giants, on the
other. The other lines in the diagram indicate the expected locations
of the stellar populations that might be present in the CMD; see text
for details on how these were calculated. The numbers next to the
BSS and SMC tracks indicate the masses of stars at the corresponding
location on these tracks; the numbers next to the WD track indicate
the WD temperature at the corresponding location on the track. 
Finally, the letters enclosed by open circles mark the positions of
field CVs if they were observed at the distance and reddening of 
47~Tuc. The letters indicate the following sources: WZ~=~WZ~Sge;
U~=~U~Gem; SS~=~SS~Cyg; VW~=~VW~Hyi; UX~=~UX~UMa; GK~=~GK~Per; AM =
AM~Her; DQ~=~DQ~Her.\label{fig3}}

\clearpage
\newpage
\pagebreak

\begin{figure}
\epsscale{0.60}
\figurenum{1}
\plotone{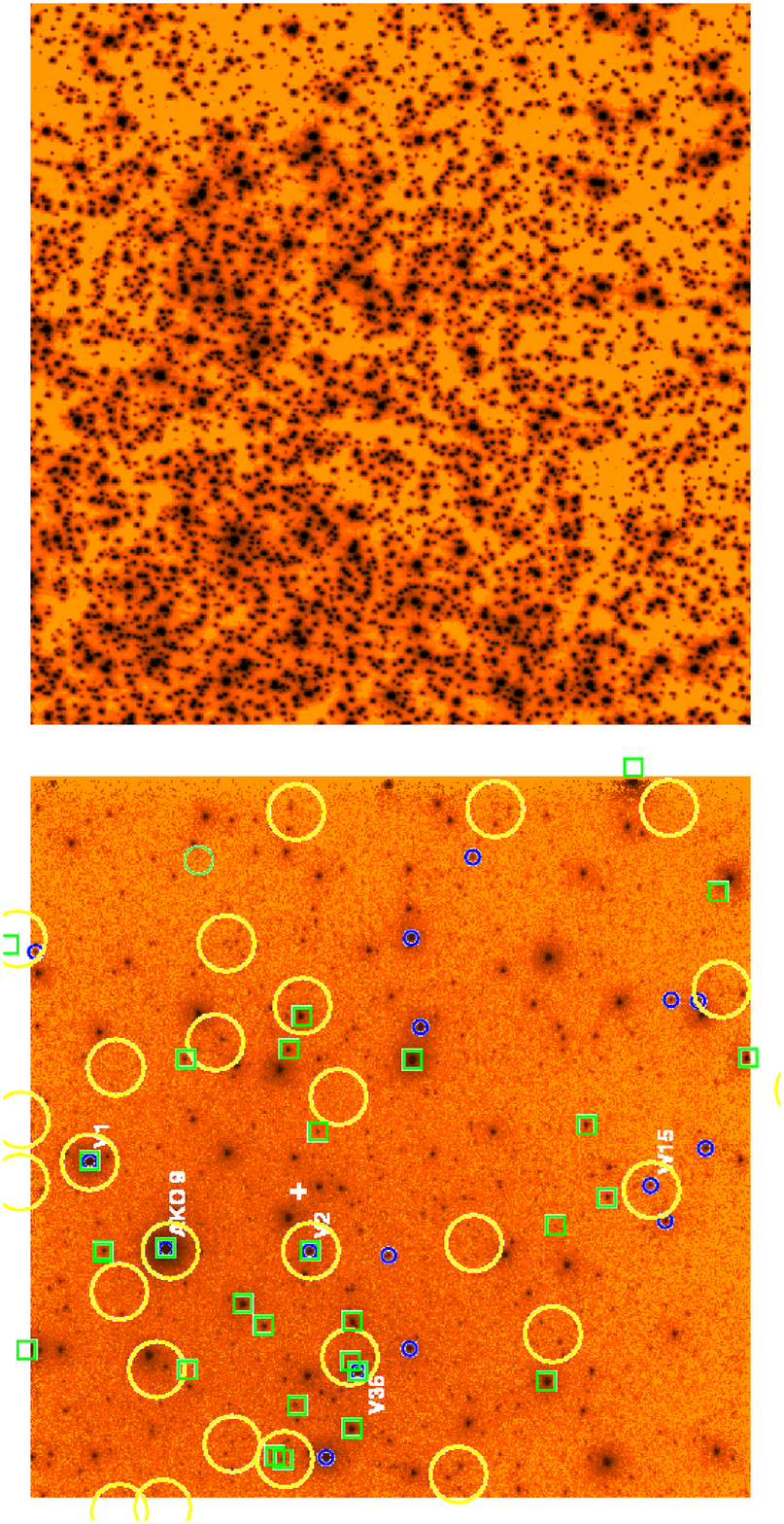}
\caption{}
\end{figure}

\clearpage
\newpage
\pagebreak

\begin{figure}
\epsscale{1.0}
\figurenum{2}
\plotone{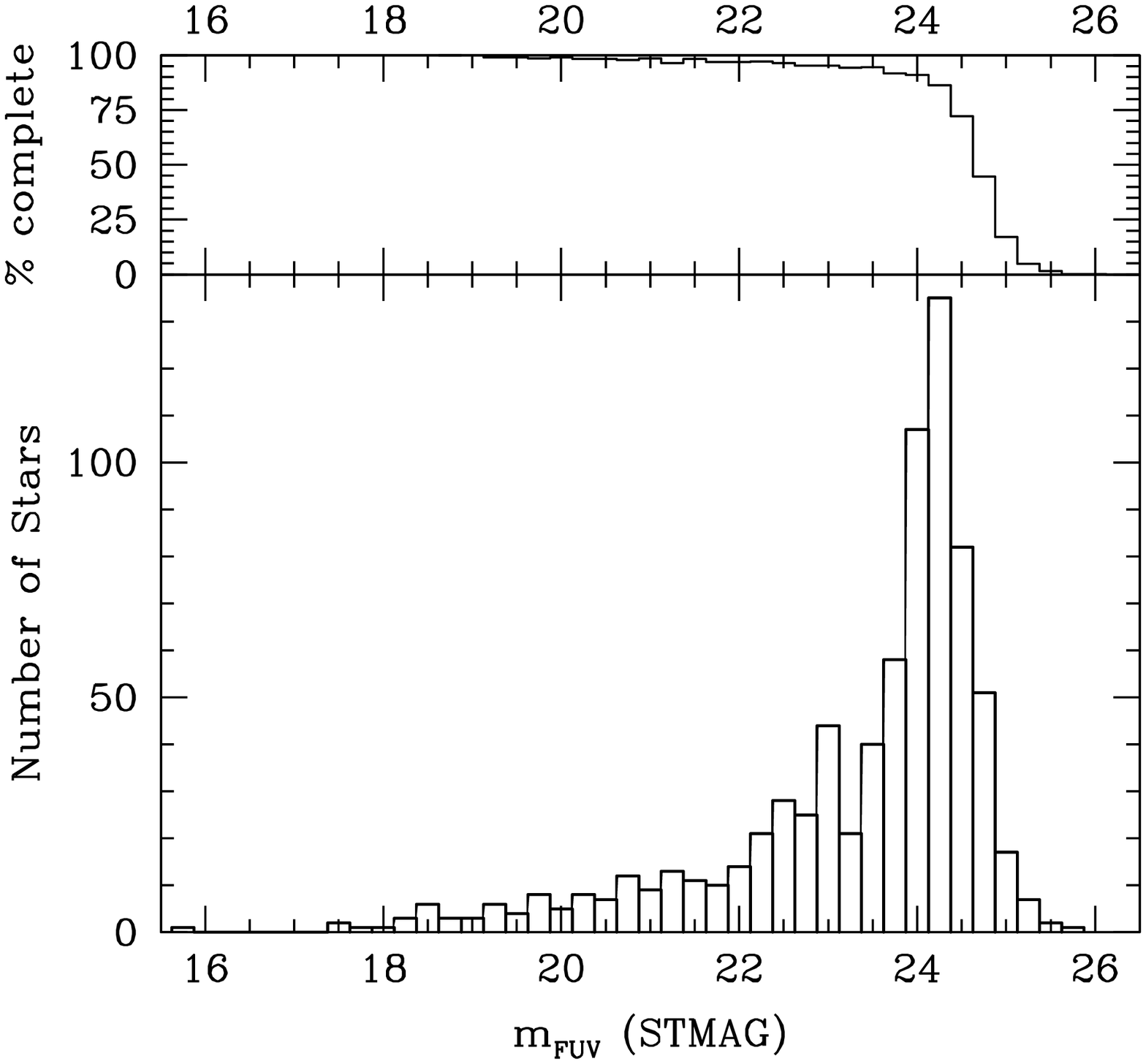}
\caption{}
\end{figure}

\clearpage
\newpage
\pagebreak

\begin{figure}
\epsscale{1.0}
\figurenum{3}
\plotone{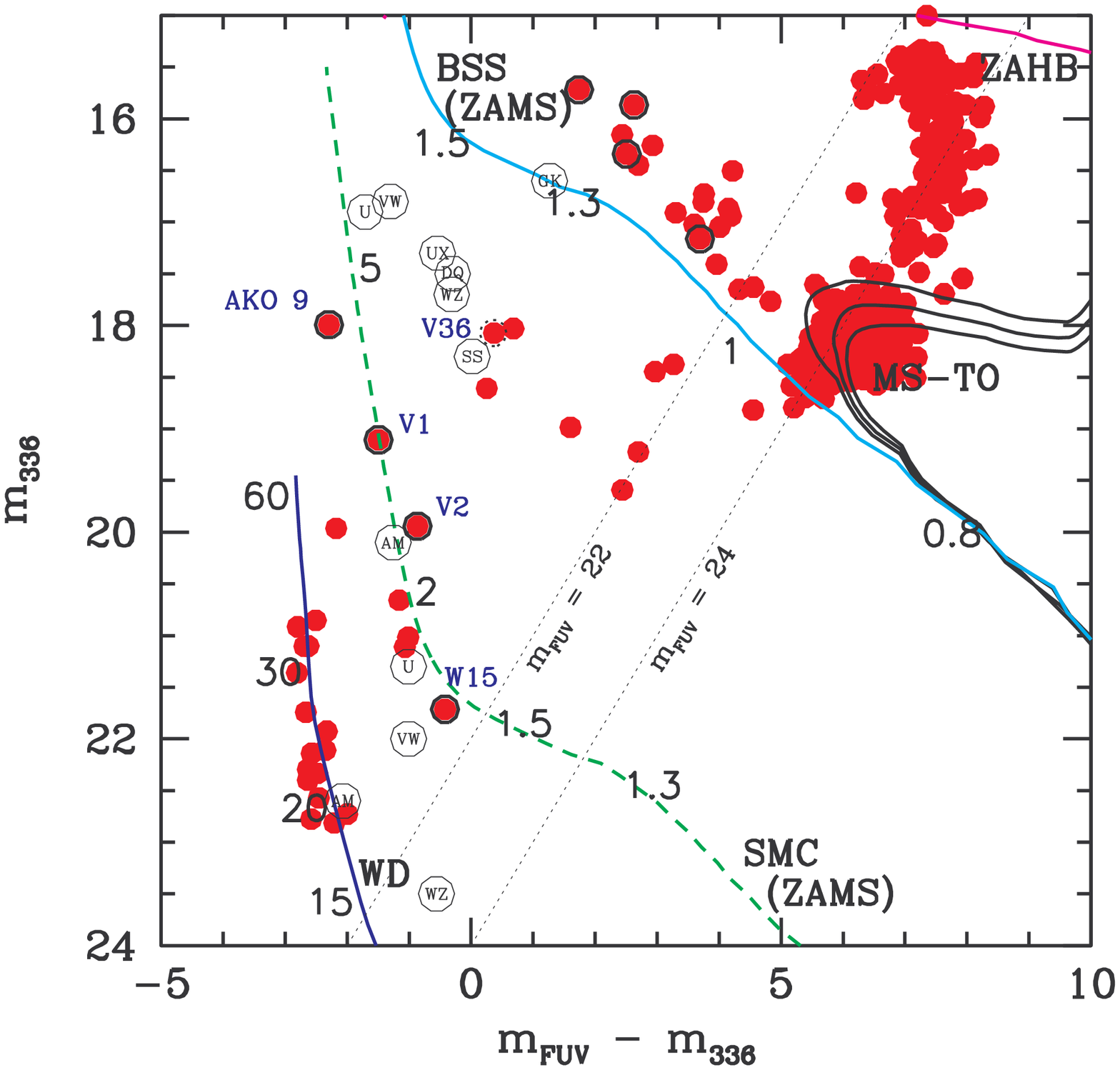}
\caption{}
\end{figure}


\end{document}